\documentclass[aps,prd,showpacs]{revtex4}
\usepackage{amsmath,amscd,amssymb}
\begin{document}

\title{Symmetries of distributional domain wall geometries}

\author{Nelson Pantoja and Alberto Sanoja
\\{\it Centro de Astrof\'{\i}sica Te\'orica,
Universidad de Los Andes, M\'erida 5101, Venezuela}}

\begin{abstract}

Generalizing the Lie derivative of smooth tensor fields to
distribution-valued tensors, we examine the Killing symmetries and
the collineations of the curvature tensors of some distributional
domain wall geometries. The chosen geometries are rigorously the
distributional thin wall limit of self gravitating scalar field
configurations representing thick domain walls and the permanence
and/or the rising of symmetries in the limit process is studied.
We show that, for all the thin wall spacetimes considered, the
symmetries of the distributional curvature tensors turns out to be
the Killing symmetries of the pullback of the metric tensor to the
surface where the singular part of these tensors is supported.
Remarkably enough, for the non-reflection symmetric domain wall
studied, these Killing symmetries are not necessarily symmetries
of the ambient spacetime on both sides of the wall.

\end{abstract}

\maketitle

\section{Introduction}

Consider a family of spacetimes
$(\mathcal{M},\,^\gamma\mathbf{g})$, where $^\gamma\mathbf{g}$ is
$C^{\infty}$ metric tensor which depends on a parameter $\gamma$.
An isometry $\psi$ on $(\mathcal{M},\,^\gamma\mathbf{g})$ is
defined to be a diffeomorphism $\psi:\mathcal{M}\rightarrow
\mathcal{M}$ for which $\psi^*\mathbf{g}=\mathbf{g}$. The
infinitesimal generator of a one-parameter group $\psi_{\lambda}$
of local isometries is the $C^{\infty}$ vector field $\mathbf{V}$
on $\mathcal{M}$ that satisfies
\begin{equation}
\mathcal{L}_{_\mathbf{V}}\,^\gamma\mathbf{g}=0,\label{KV}
\end{equation}
and $\mathbf{V}$ is called a Killing vector field on
$(\mathcal{M},\,^\gamma\mathbf{g})$ relative to this group. To
every one-parameter family of Killing symmetries there is an
associated conserved quantity along the geodesics of the spacetime
and these conserved quantities are useful for integrating the
geodesic equation \cite{wald}.

Although isometries are the most important transformations on
$(\mathcal{M},\,^\gamma\mathbf{g})$, geometric symmetries other
than Killing symmetries may also be considered. Let
$^{\gamma}\mathbf{Ric}$ be the Ricci curvature tensor of
$^\gamma\mathbf{g}$. A vector field $\mathbf{V}$ on $\mathcal{M}$
that satisfies
\begin{equation}
\mathcal{L}_{_\mathbf{V}}\,^\gamma\mathbf{Ric}=0,\label{RC}
\end{equation}
is called a Ricci collineation. It is well known that for every
vector field $\mathbf{V}$ such that (\ref{KV}) is satisfied, i.e.
for every Killing vector field, (\ref{RC}) is also satisfied and
the Ricci tensor inherits the symmetries of the metric. However,
other vector fields that are not Killing vectors may exist for
which (\ref{RC}) is satisfied and these are called proper Ricci
collineations. Since $\mathbf{Ric}$ is obtained by contracting the
Riemann curvature tensor, Ricci collineations have a natural
geometrical significance and it is believed that they can be
useful to understand the interplay between geometry and physics in
general relativity (for more details on these and other geometric
symmetries, see references \cite{katzin,coley,carot,contreras}).

Now, consider a spacetime $(\mathcal{M},\mathbf{g})$ of low
differentiability, revealing itself through a lack of smoothness
of the metric $\mathbf{g}$ and its curvature tensors. Up to what
extent the classical concept of a geometric symmetry (in the
smooth case) can be carried over to this situation? Let us assume
that the metric tensors $\mathbf{g}$ and $^{\gamma}\mathbf{g}$ are
distribution-valued tensors that satisfy
\begin{equation}
\mathbf{g}\equiv \lim_{\gamma\rightarrow0}\,^\gamma\mathbf{g}
\end{equation}
(in the sense of distributions) and that their corresponding
curvature tensors have a well defined distributional meaning.
Further, assume that the following diagram holds
\begin{equation}\label{diagram1}
\begin{CD}
    ^\gamma \mathbf{g}       @>\qquad\qquad >>
    ^\gamma \mathbf{{Riem}}  @>\qquad\qquad >>
    ^\gamma\mathbf{{Ric}}    @>\qquad\qquad >>
    ^\gamma\mathbf{{G}}\\
    @VV{\gamma\rightarrow 0}V
    @VV{\gamma\rightarrow 0}V
    @VV{\gamma\rightarrow 0}V
    @VV{\gamma\rightarrow 0}V\\
    \mathbf{g}      @>\qquad\qquad >>
    \mathbf{{Riem}} @>\qquad\qquad >>
    \mathbf{{Ric}}  @>\qquad\qquad >>
    \mathbf{{G}}\\
\end{CD}
\end{equation}
where $^\gamma\mathbf{g}$, $^\gamma\mathbf{{Riem}}$,$^\gamma
\mathbf{Ric}$, $^\gamma\mathbf{G}$ and $\mathbf{g}$,
$\mathbf{{Riem}}$, $\mathbf{Ric}$, $\mathbf{G}$ are the
distribution valued metric, Riemann, Ricci and Einstein tensors of
the smooth and the distributional geometries, respectively.
Although distributional curvatures are in general ill-defined due
to the nonlinearities of general relativity, there is a class of
distributional metrics for which the Riemann curvature tensor and
its contractions can be interpreted as distributions
\cite{geroch}. Metrics for thin shells \cite{israel} are included
into this class \cite{geroch,khorrami}. Furthermore, for such a
class an appropriate notion of convergence of metrics has been
stated which ensures the convergence of the respective curvatures
\cite{geroch}, in the sense that the diagram (\ref{diagram1})
holds.

Since the derivative of a distribution is a distribution, for a
distributional metric whose curvature tensors make sense as
distributions, it makes sense to consider their geometric
symmetries. Obviously, this situation should be considered also
from the distributional point of view. For example, equation
(\ref{RC}) contains products of the Ricci tensor and the vector
field which generates the symmetry, so that for a
distribution-valued Ricci tensor such equation restricts the
vector field to be $C^{\infty}$. With this proviso, we can
consider geometric symmetries in cases in which the curvature
tensors are zero almost everywhere (when obtained within standard
differential geometry). Furthermore, for distributional geometries
such that the above diagram holds, we can study the permanence
and/or the rising of symmetries in the limit process. Thus, from
the study of these geometries we expect to get further insight
about the nature of distributional curvatures in general
relativity.

Killing symmetries of distributional metrics have been considered
previously in reference \cite{aichelburg94}, where it is shown
that the Killing fields of the Schwarzschild metric are also
Killing of its ultra-relativistic limit, the last one being a
pp-wave with a distributional $\delta$ profile
\cite{aichelburg71}. In addition, based on the analysis of the
adjoint orbits of normal-form-preserving diffeomorphisms, Killing
symmetries of impulsive pp-waves with distributional profiles have
been analyzed \cite{aichelburg95} and the existence of non-smooth
Killing vectors put forward \cite{aichelburg96}.

In this paper, adopting a different approach for a rigorous
definition of symmetries of distributional geometries, we are
concerned with symmetries of domain wall spacetimes. Such
spacetimes have been the subject of intense investigation, after
it was realized in reference \cite{randall} that our four
dimensional universe might be a thin (codimension one)
distributional domain wall embedded in a five dimensional
spacetime. Thin wall geometries have distribution-valued curvature
tensor fields whose singular parts are proportional to a Dirac
distribution supported on the surface where the wall is localized.
All the metrics representing the possible backgrounds of an
infinitely thin domain wall \cite{vilenkin} have been found and
classified \cite{cvetic93,cvetic97}, these being joined at a
common boundary, the surface where the wall is localized,
following the Darmois-Israel formalism \cite{israel}. On the other
hand, smooth domain wall geometries can be obtained as solutions
to the coupled Einstein-scalar field system with a suitable
symmetry breaking potential $V(\phi)$
\cite{vilenkin,goetz,mukherjee,bonjour,gass}. The behavior of
gravity in some of these models has been also investigated
\cite{dewolfe,gremm,wang}. Recently, following the convergence
criteria of reference \cite{geroch}, the distributional thin wall
limit of some classes of domain wall spacetimes has been
rigourously analyzed \cite{guerrero,melfo} showing that the
diagram (\ref{diagram1}), in which $^\gamma\mathbf{g}$ and
$\mathbf{g}$ are the distribution valued metric tensors of the
thick and thin domain wall spacetimes, respectively, holds.
Specifically, we will study the symmetries of the metric and its
curvature tensors of these distributional domain wall geometries.
Domain walls have drastic gravitational effects in the ambient
space and, due to its role in brane-world models, we are
interested in their geometric symmetries besides those defining
the plane-parallel symmetry since a larger group may exists for
various particular models.

In section \ref{mathframe}, after an overview on the subject of
distribution-valued tensors, we give the definition of the Lie
derivative of a tensor distribution along a $C^{\infty}$ vector
field. In the next three sections, we examine the Killing
symmetries and the collineations of the singular Ricci and
Einstein curvature tensors associated to the distributional thin
wall limit of some thick domain wall spacetimes for which the
diagram (\ref{diagram1}) holds \cite{guerrero}.  The last section
is devoted to summarize and discuss the results.

\section{Mathematical framework}\label{mathframe}

We first recall some fundamental results about distribution-valued
tensors on a $C^{\infty}$ paracompact $n$-dimensional manifold
$\mathcal{M}$ \cite{choquet,choquet2,lichnerowicz}.

Let $\mathcal{D}_p(\mathcal{M})$ be the space of $C^{\infty}$
$p$-tensor fields on $\mathcal{M}$ with compact support endowed
with its Schwartz topology, i.e., the space of test $p$-tensor
fields. The space of $p$-cotensor distributions on $\mathcal{M}$,
$\mathcal{D}^{*'}_p(\mathcal{M})$, is defined as the dual of
$\mathcal{D}_p(\mathcal{M})$. Now, in order to keep things simple,
we endow the $C^{\infty}$ paracompact manifold $\mathcal{M}$ with
a $C^{\infty}$ metric $\mathbf{\eta}$. However, it should be noted
that tensor distributions and their derivatives can be described
without assuming the presence of a metric \cite{dray} and that, by
using de Rham currents \cite{rham} and replacing test tensors by
test $n$-forms, the introduction of a volume form can also be
avoided. Here we shall follow the approach of \cite{choquet}.

Let ${\bf U}$ be a test $p$-tensor field on ${\cal M}$. The
identification of a locally integrable $p$-cotensor field
$\mathbf{T}$ with a distribution-valued tensor is defined by
\begin{equation}
{\bf T}[{\bf U}]\equiv \int_{\cal M} {\bf T}\cdot{\bf
U}\,\boldsymbol{\omega}_{\eta}, \label{Tdist}
\end{equation}
where ${\bf T}\cdot{\bf U}$ denotes the scalar product of
$\mathbf{T}$ and ${\bf U}$, and $\boldsymbol{\omega}_{\eta}$ is
the volume element of $\boldsymbol{\eta}$. Since (\ref{Tdist}) is
the integral of an $n$-form with compact support, we have
\begin{equation}
{\bf T}[{\bf U}]\equiv \int_{\varphi({\cal M})} T_{i_1\ldots
i_p}U^{i_1\ldots i_p} |\det {\bf \eta}|^{\frac{1}{2}} dx^1\ldots
dx^n \label{Tdist2}
\end{equation}
in the domain $\varphi({\cal M})$ of the chart $(x^1,\ldots,x^n)$.
Obviously, (\ref{Tdist2}) is independent of the choice of
coordinate system covering the corresponding domain.

Next, let us define the Lie derivative of  a tensor field in the
sense of distributions. Let $\mathbf{V}$ be a $C^{\infty}$ vector
field and $\mathbf{T}$ be a $C^1$ $p$-cotensor field on
$\mathcal{M}$. The Lie derivative of $\mathbf{T}$ is the
$p$-cotensor field $\mathcal{L}_{_\mathbf{V}} \mathbf{T}$ such
that, for every test $p$-tensor field ${\bf U}$
\begin{equation}
\mathcal{L}_{_\mathbf{V}} \mathbf{T}[{\bf U}] \equiv \int_{\cal M}
\mathcal{L}_{_\mathbf{V}} {\bf T}\cdot{\bf
U}\,\boldsymbol{\omega}_{\eta}= \int_{\cal M}
\mathcal{L}_{_\mathbf{V}} ({\bf T}\cdot{\bf
U}\,\boldsymbol{\omega}_{\eta}) - \int_{\cal M} {\bf
T}\cdot\mathcal{L}_{_\mathbf{V}} ({\bf
U}\,\boldsymbol{\omega}_{\eta})\,.
\end{equation}
Now, since ${\bf T}\cdot{\bf U}\, \boldsymbol{\omega}_{\eta}$ is
an $n$-form with compact support, we have
\begin{equation}
\int_\mathcal{M} \mathcal{L}_{_\mathbf{V}}({\bf T}\cdot{\bf U}\,
\boldsymbol{\omega}_{\eta})= \int_\mathcal{M}
{\mathbf{d\,i}}_{_\mathbf{V}}({\bf T}\cdot{\bf U}\,
\boldsymbol{\omega}_{\eta})= \int_{\partial
\mathcal{M}}{\mathbf{i}}_{_\mathbf{V}}({\bf T}\cdot{\bf U}\,
\boldsymbol{\omega}_{\eta})= 0,
\end{equation}
where ${\mathbf{i}}_{_\mathbf{V}}$ denotes interior product, in
the last step we have used Stokes' theorem and the surface term
vanishes because $\mathbf{U}$ has compact support. Therefore
\begin{equation}
\mathcal{L}_{_\mathbf{V}} \mathbf{T}[{\bf U}]= -\int_\mathcal{M}
{\bf T} \cdot\mathcal{L}_{_\mathbf{V}}({\bf U}\,
\boldsymbol{\omega}_{\eta})=-\int_\mathcal{M} ({\bf T}
\cdot\mathcal{L}_{_\mathbf{V}}{\bf U}) \boldsymbol{\omega}_{\eta}
+ (\mathbf{T}\cdot\mathbf{U})
\mathcal{L}_{_\mathbf{V}}\boldsymbol{\omega}_{\eta}\,.\label{rhs}
\end{equation}
On the other hand, we have
\begin{equation}
\mathcal{L}_{_\mathbf{V}}\boldsymbol{\omega}_{\eta}=
(\nabla\cdot\mathbf{V})\boldsymbol{\omega}_{\eta}\label{rhs-2t}
\end{equation}
where $\nabla$ is the derivative in $\boldsymbol{\eta}$ and
\begin{equation}
\nabla\cdot\mathbf{V}\equiv\nabla_j V^j.\label{rhs-3}
\end{equation}
It follows that
\begin{equation}
\mathcal{L}_{_\mathbf{V}} \mathbf{T}[{\bf U}]= -\int_\mathcal{M}
{\bf T} \cdot(\mathcal{L}_{_\mathbf{V}}{\bf U} +
\mathbf{U}(\nabla\cdot\mathbf{V}))\boldsymbol{\omega}_{\eta}\,.\label{liederivsmooth}
\end{equation}
Therefore, to have a definition which coincides with the usual one
when $\mathbf{T}$ is $C^1$, we define for $\mathbf{T}$ an
arbitrary $p$-cotensor distribution and $\mathbf{V}$ a
$C^{\infty}$ vector field on $\mathcal{M}$ the Lie derivative of
$\mathbf{T}$ as the $p$-cotensor distribution given by
\begin{equation}
\mathcal{L}_{_\mathbf{V}} \mathbf{T}[{\bf U}]\equiv -
\mathbf{T}[\mathcal{L}_{_\mathbf{V}} \mathbf{U} +
(\nabla\cdot\mathbf{V})\mathbf{U}].\label{liederiv}
\end{equation}
Note that (\ref{liederiv}) makes sense since both
$\mathcal{L}_{_\mathbf{V}} \mathbf{U}$ and
$(\nabla\cdot\mathbf{V})\,\mathbf{U}$ are test $p$-tensor fields
if $\mathbf{V}$ is a $C^{\infty}$ vector field.

In the next sections, (\ref{liederiv}) will be used to define and
compute the Lie derivative of a distribution-valued metric tensor
$\mathbf{g}$ with $\mathbf{V}$ a $C^{\infty}$ vector field that
generates a one parameter group of isometries of the spacetime
$(\mathcal{M},\mathbf{g})$. We also make use of (\ref{liederiv})
to define and compute the Ricci and Einstein collineations of the
distributional geometry associated to these spacetimes. Since all
the distribution-valued metric tensors that we shall consider are
regular metrics in the sense of reference \cite{geroch}, let us
recall their definition. Suppose that $({\cal M},\mathbf{g})$ are
given such that
\begin{enumerate}
\item $\mathbf{g}$ and $(\mathbf{g}^{-1})$ exist everywhere and
are locally bounded, \item the first derivative $\nabla
\mathbf{g}$ (in the sense of distributions) of $\mathbf{g}$ in
some smooth derivative operator $\nabla$ exists and is locally
square-integrable, i.e. the outer product of the derivative with
itself is locally integrable.
\end{enumerate}
Following reference \cite{geroch}, these are the minimal
conditions for the Riemann curvature tensor to be definable as a
distribution by the usual coordinate formula and we shall say that
$\mathbf{g}$ is a regular metric. Since for this class of metrics
the outer product of any number of metrics and inverse metrics
with the Riemann curvature tensor can be interpreted as
distributions, the Ricci and Einstein curvature tensors of a
regular metric make sense as distributions. Finally, it should be
stressed that for such class an appropriate notion of convergence
of metrics exists which ensures the convergence of the respective
curvatures \cite{geroch}, in the sense that the diagram
(\ref{diagram1}) holds (For details, see reference \cite{geroch}).

\section{A domain wall with a de Sitter
expansion}\label{Gspacetime}

Consider the spacetime $({\mathbb{R}}^4,\,^\gamma\mathbf{g})$
where the metric tensor $^\gamma\mathbf{g}$, in a particular
coordinate basis, is given by
\begin{equation}
^\gamma\mathbf{g}= \cosh^{-2\gamma}\frac{\beta
x}{\gamma}\left(-\mathbf{d}t\, \mathbf{d}t + \mathbf{d}x\,
\mathbf{d}x + e^{2\beta t}(\mathbf{d}y\, \mathbf{d}y +
\mathbf{d}z\, \mathbf{d}z)\right),\label{goetz}
\end{equation}
where $\beta$ and $\gamma$ are constants, with $0<\gamma < 1$.
This spacetime is generated by a thick domain wall, i.e. a
solution to the coupled Einstein-scalar field equations
\begin{equation}
^\gamma\mathbf{G}+ \,^{\gamma}\mathbf{g}\,\Lambda \equiv
\,^\gamma\mathbf{Ric} -\frac{1}{2}\,^\gamma\mathbf{g}\,
^\gamma{\mathrm{R}} + \,^{\gamma}\mathbf{g}\,\Lambda= 8\pi
\left[{\mathbf{D}}\phi\, {\mathbf{D}}\phi- \mathbf{g}\, \left(
\frac{1}{2}({\mathbf{D}}\phi \,| \,{\mathbf{D}}\phi) +
V(\phi)\right)\right]\label{eqEinstein}
\end{equation}
and
\begin{equation}
\Box \phi - \frac{d}{d\phi} V(\phi)=0,\label{eqScalar}
\end{equation}
where $\Box\equiv \Delta\mathbf{D}+ \mathbf{D}\Delta$, with
$\Lambda=0$, and
\begin{equation}
\phi= \phi_0 \tan^{-1}(\sinh(\beta x/\gamma),\qquad \phi_0\equiv
\sqrt{\gamma(1-\gamma)/4\pi},
\end{equation}
\begin{equation}
V(\phi)= \frac{1}{8\pi}\beta^2 (2+\frac{1}{\gamma})
\cos^{2(1-\gamma)}(\phi/\phi_0).
\end{equation}
The scalar field takes values $\pm\pi\phi_0/2$ at $x\pm\infty$
corresponding to two consecutive minima of the potential and
interpolates smoothly between these values
\cite{goetz,mukherjee,gass}, with $\gamma$ playing the role of the
wall's thickness \cite{guerrero}. The five-dimensional analogue of
this geometry, considered as a thick brane-world model, has been
studied in reference \cite{wang}.

Note that (\ref{goetz}) is $C^{\infty}$ as are also all its
curvature tensor fields. In particular, for the Ricci and Einstein
tensor fields we obtain
\begin{equation}
^{\gamma}\mathbf{Ric}= -(2+\frac{1}{\gamma})\,\beta^2
\cosh^{-2}\frac{\beta x}{\gamma}\left(-\mathbf{d}t\, \mathbf{d}t +
\frac{3}{1+2\gamma}\,\mathbf{d}x\, \mathbf{d}x + e^{2\beta
t}(\mathbf{d}y\, \mathbf{d}y + \mathbf{d}z\,
\mathbf{d}z)\right)\label{Rgoetz}
\end{equation}
and
\begin{equation}
^{\gamma}\mathbf{G}= -(1+\frac{2}{\gamma})\,\beta^2
\cosh^{-2}\frac{\beta x}{\gamma}\left(-\mathbf{d}t\, \mathbf{d}t +
\frac{3\gamma}{2+\gamma}\,\mathbf{d}x\, \mathbf{d}x + e^{2\beta
t}(\mathbf{d}y\, \mathbf{d}y + \mathbf{d}z\,
\mathbf{d}z)\right).\label{Ggoetz}
\end{equation}

Now, from (\ref{KV}) and (\ref{goetz}), we obtain six linearly
independent Killing vector fields on
$({\mathbb{R}}^4,\,^\gamma\mathbf{g})$ given by
\begin{equation}
\mathbf{V}_1= \boldsymbol{\partial}_y,\qquad \mathbf{V}_2=
\boldsymbol{\partial}_z,\qquad \mathbf{V}_3=
z\boldsymbol{\partial}_y
-y\boldsymbol{\partial}_z,\label{goetzcoll1}
\end{equation}
\begin{equation}
\mathbf{V}_4=
\boldsymbol{\partial}_t-\beta(y\boldsymbol{\partial}_y+z\boldsymbol{\partial}_z),
\label{goetzcoll2}
\end{equation}
\begin{equation}
\mathbf{V}_5= 2\beta y\mathbf{V}_4 + \left(\beta^2 (y^2+z^2) -
e^{-2\beta t}\right)\boldsymbol{\partial}_y,\label{goetzcoll3}
\end{equation}
\begin{equation}
\mathbf{V}_6= 2\beta z\mathbf{V}_4 + \left(\beta^2 (y^2+z^2) -
e^{-2\beta t}\right)\boldsymbol{\partial}_z\, .\label{goetzcoll4}
\end{equation}
The Killing vectors (\ref{goetzcoll1}) are generic to the plane
parallel symmetry, two spatial translations $\mathbf{V}_1$,
$\mathbf{V}_2$ and one spatial rotation $\mathbf{V}_3$. The six
vector fields (\ref{goetzcoll1}-\ref{goetzcoll4}) are also Killing
vectors on the $(2+1)$-dimensional de Sitter spacetime
$(\mathbb{R}^3,\bar{\mathbf{g}})$ with
\begin{equation}
\bar{\mathbf{g}}= -\mathbf{d}t\, \mathbf{d}t + e^{2\beta
t}(\mathbf{d}y\, \mathbf{d}y + \mathbf{d}z\,
\mathbf{d}z),\label{de Sitter}
\end{equation}
where $\mathbf{V}_4$ is a quasi-time translation and
$\mathbf{V}_5$, $\mathbf{V}_6$ are quasi-Lorentz rotations. Note
that (\ref{goetzcoll1}-\ref{goetzcoll4}) are all independent of
the thickness $\gamma$ of the wall. It is straightforward to show
that for (\ref{goetzcoll1}-\ref{goetzcoll4}) we have
\begin{equation}
\mathcal{L}_{_\mathbf{V}}\,^\gamma\mathbf{Ric}=0,\qquad
\mathcal{L}_{_\mathbf{V}}\,^\gamma\mathbf{G}=0. \label{R-E coll}
\end{equation}
One can also show that there are no other vector fields for which
(\ref{R-E coll}) is satisfied. Hence, the Ricci (\ref{Rgoetz}) and
Einstein (\ref{Ggoetz}) curvature tensors only admit improper
collineations.

Next, consider the Lie derivative of (\ref{goetz}) in the sense of
distributions as given by (\ref{liederiv}). Let
$\boldsymbol{\eta}$ be the ordinary four-dimensional Minkowski
metric tensor in cartesian coordinates
\begin{equation}
\boldsymbol{\eta}= -\mathbf{d}t\, \mathbf{d}t + \mathbf{d}x\,
\mathbf{d}x + \mathbf{d}y\ \mathbf{d}y + \mathbf{d}z\,
\mathbf{d}z,\label{minkowski}
\end{equation}
and let $\nabla$ be the derivative operator in
$\boldsymbol{\eta}$. Let $\mathbf{U}$ be a test tensor on
$\mathbb{R}^4$. From (\ref{liederiv}), it is straightforward to
verify that for all Killing vectors
(\ref{goetzcoll1}-\ref{goetzcoll4}), we have
\begin{equation}
\mathcal{L}_{_\mathbf{V}} \,\mathbf{^{\gamma}g}[{\bf U}]=0\,
,\label{liederivgoetz}
\end{equation}
as expected. Indeed, it is to make such things true that the Lie
derivative is defined as in (\ref{liederiv}). For the sake of
completeness, let us to show explicitly the computation of this
derivative for the Killing vector given by (\ref{goetzcoll2}). We
have
\begin{eqnarray}
\mathcal{L}_{_\mathbf{V_4}}\,^{\gamma}\mathbf{g}[{\bf U}]&\equiv&
- ^{\gamma}\mathbf{g}[\mathcal{L}_{_\mathbf{V_4}} \mathbf{U} +
(\nabla\cdot\mathbf{V_4})\mathbf{U}]\nonumber\\&=&
-\int_{\mathbb{R}^4}\, ^{\gamma}{\bf g}
\cdot(\mathcal{L}_{_\mathbf{V_4}}{\bf U} +
\mathbf{U}(\nabla\cdot\mathbf{V_4}))\boldsymbol{\omega}_{\eta}\nonumber\\&=&
-\int_{\mathbb{R}^4}\cosh^{-2\gamma}\frac{\beta
x}{\gamma}\left((\partial_t - \beta y\partial_y - \beta
z\partial_z - 2\beta)(-U^{tt} + U^{xx}) \right. \nonumber \\ &&
\hspace{3cm}\left. + e^{2\beta t}(\partial_t - \beta y\partial_y -
\beta z\partial_z)(U^{yy} + U^{zz}))\right)dt\,dx\,dy\,dz
\nonumber\\&=&0,
\end{eqnarray}
where in last step we have integrated by parts and used the fact
that $\mathbf{U}$ is of compact support. For all the Killing
vector fields (\ref{goetzcoll1}-\ref{goetzcoll4}), analogous
computations show that
\begin{equation}
\mathcal{L}_{_\mathbf{V}}\,^{\gamma} \mathbf{Ric}[{\bf U}]=0,
\qquad \mathcal{L}_{_\mathbf{V}}\,^{\gamma} \mathbf{G}[{\bf U}]=0,
\end{equation}
as expected.

We turn now to consider the $\gamma\rightarrow 0$ limit of this
geometry. In reference \cite{guerrero}, it has been proved that
(\ref{goetz}) provides a sequence of metrics that satisfies the
convergence conditions of \cite{geroch} such that the limit of the
Riemann curvature tensor exists and is the Riemann tensor of the
limit metric. The same holds for the other curvature tensors. We
have
\begin{equation}
\mathbf{g}\equiv \lim_{\gamma\rightarrow0}\,^\gamma\mathbf{g}=
(\Theta^-_x \,e^{2\beta x} + \Theta^+_x \,e^{-2\beta
x})\left(-\mathbf{d}t\, \mathbf{d}t + \mathbf{d}x\, \mathbf{d}x +
e^{2\beta t}(\mathbf{d}y\, \mathbf{d}y + \mathbf{d}z\,
\mathbf{d}z)\right),\label{limgoetz}
\end{equation}
\begin{equation}
\mathbf{Ric}\equiv \lim_{\gamma\rightarrow0}\,^\gamma\mathbf{Ric}=
2\beta\,\delta_0\left(\mathbf{d}t\, \mathbf{d}t - 3\mathbf{d}x\,
\mathbf{d}x - e^{2\beta t}(\mathbf{d}y\, \mathbf{d}y +
\mathbf{d}z\, \mathbf{d}z)\right),\label{limRgoetz}
\end{equation}
\begin{equation}
\mathbf{G}\equiv \lim_{\gamma\rightarrow0}\,^\gamma\mathbf{G}=
4\beta\,\delta_0\left(\mathbf{d}t\, \mathbf{d}t - e^{2\beta
t}(\mathbf{d}y\, \mathbf{d}y + \mathbf{d}z\,
\mathbf{d}z)\right),\label{limGgoetz}
\end{equation}
where $\Theta^-_x$ and $\Theta^+_x$ are the Heaviside
distributions with support on $x<0$ and $x>0$ respectively and
$\delta_0$ is the Dirac measure with support on the surface $x=0$.
Note that $\mathbf{g}$ is piecewise smooth and that the pullback
of $\mathbf{g}$ to the surface $x=0$ is the same from both sides.
Note also that this pullback is given by (\ref{de Sitter}).
Indeed, the above expressions should be understood in the sense of
distributions. Thus, $\mathbf{g}\equiv
\lim_{\gamma\rightarrow0}\,^\gamma\mathbf{g}$ means
\begin{equation}
\mathbf{g}[\mathbf{U}]\equiv
\lim_{\gamma\rightarrow0}\,^\gamma\mathbf{g}[\mathbf{U}].
\end{equation}
In fact, $^\gamma\mathbf{g}, \,(^\gamma\mathbf{g})^{-1}$ and
$\nabla\,^\gamma\mathbf{g}$ converge locally (in square integral)
to $\mathbf{g}, \,(\mathbf{g})^{-1}$ and $\nabla\mathbf{g}$
respectively and $\mathbf{Ric}$ and $\mathbf{G}$ are the
distribution-valued Ricci and Einstein curvatures of $\mathbf{g}$
\cite{guerrero}. It follows that the diagram (\ref{diagram1}), in
which $^\gamma\mathbf{g}$ and $\mathbf{g}$ are the distribution
valued metric tensors of the thick and thin domain wall
spacetimes, respectively, holds in the sense of distributions.

Now, as follows from the distributional convergence
(\ref{diagram1}) proved in reference \cite{guerrero} and the fact
that the vectors (\ref{goetzcoll1}-\ref{goetzcoll4}) are all
smooth vector fields independent of the wall's thickness, these
are also Killing vectors of $(\mathbb{R}^4,\mathbf{g})$ in the
sense that $\mathcal{L}_{_\mathbf{V}}\,\mathbf{g}$ is the zero
distribution on $\mathbb{R}^4$. As an example, consider the
Killing vector given by (\ref{goetzcoll2}). For (\ref{limgoetz})
we have
\begin{eqnarray}
\mathcal{L}_{_\mathbf{V_4}}\,\mathbf{g}[{\bf U}]&\equiv& -
\mathbf{g}[\mathcal{L}_{_\mathbf{V_4}} \mathbf{U} +
(\nabla\cdot\mathbf{V_4})\mathbf{U}]\nonumber\\&=&
-\int_{\mathbb{R}^4}\,{\bf g}
\cdot(\mathcal{L}_{_\mathbf{V_4}}{\bf U} +
\mathbf{U}(\nabla\cdot\mathbf{V_4}))\boldsymbol{\omega}_{\eta}\nonumber\\&=&
-\int_{x<0}e^{2\beta x}\left((\partial_t - \beta y\partial_y -
\beta z\partial_z - 2\beta)(-U^{tt} + U^{xx}) \right. \nonumber \\
&& \hspace{3cm}\left. + e^{2\beta t}(\partial_t - \beta
y\partial_y - \beta z\partial_z)(U^{yy} +
U^{zz}))\right)dt\,dx\,dy\,dz \nonumber\\& \, &
-\int_{x>0}e^{-2\beta x}\left((\partial_t - \beta y\partial_y -
\beta z\partial_z - 2\beta)(-U^{tt} + U^{xx}) \right. \nonumber \\
&& \hspace{3cm}\left. + e^{2\beta t}(\partial_t - \beta
y\partial_y - \beta z\partial_z)(U^{yy} +
U^{zz}))\right)dt\,dx\,dy\,dz \nonumber\\&=& 0,
\end{eqnarray}
where we have integrated by parts and used the fact that
$\mathbf{U}$ is of compact support.

Indeed, (\ref{goetzcoll1}-\ref{goetzcoll4}) are also improper
Ricci collineations. For (\ref{limRgoetz}) and (\ref{goetzcoll2})
we have
\begin{eqnarray}
\mathcal{L}_{_\mathbf{V_4}}\,\mathbf{Ric}[{\bf U}]&\equiv& -
\mathbf{Ric}[\mathcal{L}_{_\mathbf{V_4}} \mathbf{U} +
(\nabla\cdot\mathbf{V_4})\mathbf{U}]\nonumber\\&=&
-\beta\int_{x=0}\left((\partial_t - \beta y\partial_y - \beta
z\partial_z - 2\beta)(2U^{tt} - 3U^{xx}) \right. \nonumber
\\ && \hspace{3cm}\left. - 2e^{2\beta t}(\partial_t - \beta
y\partial_y - \beta z\partial_z)(U^{yy} +
U^{zz}))\right)dt\,dy\,dz \nonumber\\&=&0,
\end{eqnarray}
where in last step we have integrated by parts and used again the
fact that $\mathbf{U}$ is of compact support. Analogous
computations show that
\begin{equation}
\mathcal{L}_{_\mathbf{V}}\,\mathbf{Ric}[\mathbf{U}]=0,\qquad
\mathcal{L}_{_\mathbf{V}}\,\mathbf{G}[\mathbf{U}]=0, \label{limR-E
coll}
\end{equation}
for all the Killing symmetries of (\ref{goetz}), a result that
obviously can not be proved outside the distributional setting. It
follows that for this spacetime the diagram
\begin{equation}\label{diagram2}
\begin{CD}
    \mathcal{L}_{_\mathbf{V}}\,^\gamma \mathbf{g}=0       @>\qquad\qquad >>
    \mathcal{L}_{_\mathbf{V}}\,^\gamma\mathbf{{Ric}}=0    @>\qquad\qquad >>
    \mathcal{L}_{_\mathbf{V}}\,^\gamma\mathbf{{G}}=0\\
    @VV{\gamma\rightarrow 0}V
    @VV{\gamma\rightarrow 0}V
    @VV{\gamma\rightarrow 0}V\\
    \mathcal{L}_{_\mathbf{V}}\,\mathbf{g}=0      @>\qquad\qquad >>
    \mathcal{L}_{_\mathbf{V}}\,\mathbf{{Ric}}=0  @>\qquad\qquad >>
    \mathcal{L}_{_\mathbf{V}}\,\mathbf{{G}}=0\\
\end{CD}
\end{equation}
holds in the sense of distributions. Besides
(\ref{goetzcoll1}-\ref{goetzcoll4}), there is no other
$C^{\infty}$ vector field for which (\ref{limR-E coll}) are
satisfied. It should be noted that both diagrams, (\ref{diagram1})
and (\ref{diagram2}), hold under $\gamma$-dependent smooth
diffeomorphisms, whenever in the limit $\gamma\rightarrow 0$ these
remain bounded in order to avoid different identifications of
points in the manifold under these diffeomophisms.

\section{A domain wall embedded in an anti-de Sitter
spacetime}\label{GMPspacetime}

 Consider the spacetime
$({\mathbb{R}}^4,\,^\gamma\mathbf{g})$ where the metric tensor is
given by
\begin{equation}
^\gamma\mathbf{g}= \cosh^{-2\gamma}\frac{\beta
x}{\gamma}\left(-\mathbf{d}t\, \mathbf{d}t + \mathbf{d}y\,
\mathbf{d}y + \mathbf{d}z\, \mathbf{d}z\right) + \mathbf{d}x\,
\mathbf{d}x,\label{gmp}
\end{equation}
with $\beta$ and $\gamma$ constants and $\gamma>0$. This is a
thick domain wall spacetime, solution to the coupled
Einstein-scalar field equations (\ref{eqEinstein}-\ref{eqScalar})
with $\Lambda=-3\beta^2$ and
\begin{equation}
\phi= \phi_0 \tan^{-1}(\sinh(\beta x/\gamma)),\qquad \phi_0\equiv
\sqrt{\gamma /4\pi}
\end{equation}
and
\begin{equation}
V(\phi)= \frac{1}{8\pi}\beta^2 \left(3 + \frac{1}{\gamma}\right)\,
\cos^{2}(\phi/\phi_0).
\end{equation}
This spacetime behaves asimptotically (i.e. far away of the wall)
as an anti de Sitter spacetime \cite{guerrero} and its five
dimensional analogue provides a thick domain wall version
\cite{gremm} of the original Randall-Sundrum scenario
\cite{randall}. Here, $\gamma$ plays the role of the wall's
thickness and the distributional $\gamma \rightarrow 0$ limit of
this geometry has been analyzed in reference \cite{guerrero}.

There are six Killing vector fields on
$({\mathbb{R}}^4,\,^\gamma\mathbf{g})$, given by
\begin{equation}
\qquad \mathbf{V}_1= \boldsymbol{\partial}_y,\qquad \mathbf{V}_2=
\boldsymbol{\partial}_z,\qquad \mathbf{V}_3=
z\boldsymbol{\partial}_y -y\boldsymbol{\partial}_z,
\label{gmpcoll1}
\end{equation}
\begin{equation}
\mathbf{V}_4= \boldsymbol{\partial}_t,\qquad \mathbf{V}_5=
t\boldsymbol{\partial}_y + y\boldsymbol{\partial}_t,\qquad
\mathbf{V}_6 = z\boldsymbol{\partial}_t +t\boldsymbol{\partial}_z,
\label{gmpcoll2}
\end{equation}
where $\mathbf{V}_4$ is a time translation and
$\mathbf{V}_5,\,\mathbf{V}_6$ are Lorentz rotations.

The Ricci and Einstein tensor fields of (\ref{gmp}) are given by
\begin{equation}
^{\gamma}\mathbf{Ric}= \Lambda\tanh^2\frac{\beta
x}{\gamma}\,^{\gamma}\mathbf{g} + \frac{\beta^2}{\gamma}
\cosh^{-2}\frac{\beta x}{\gamma}\left( \cosh^{-2\gamma}\frac{\beta
x}{\gamma}\left(-\mathbf{d}t\, \mathbf{d}t +
 \mathbf{d}y\, \mathbf{d}y + \mathbf{d}z\,
\mathbf{d}z\right) + 3\mathbf{d}x\,
\mathbf{d}x\right),\label{Rgmp}
\end{equation}
\begin{equation}
^{\gamma}\mathbf{G}= -\Lambda\tanh^2\frac{\beta
x}{\gamma}\,^{\gamma}\mathbf{g} - \frac{2\beta^2}{\gamma}
\cosh^{-2(\gamma + 1)}\frac{\beta x}{\gamma}\left(-\mathbf{d}t\,
\mathbf{d}t +
 \mathbf{d}y\, \mathbf{d}y + \mathbf{d}z\,
\mathbf{d}z\right).\label{Ggmp}
\end{equation}
The distributional thin wall limits of (\ref{gmp}) and
(\ref{Rgmp},\ref{Ggmp}) are given by
\begin{equation}
\mathbf{g}\equiv \lim_{\gamma\rightarrow0}\,^\gamma\mathbf{g}=
\left(\Theta^-_x e^{2\beta x} + \Theta^+_x e^{-2\beta
x}\right)(-\mathbf{d}t\, \mathbf{d}t + \mathbf{d}y\, \mathbf{d}y +
\mathbf{d}z\, \mathbf{d}z) + \mathbf{d}x\, \mathbf{d}x\,
,\label{limgmp}
\end{equation}
\begin{equation}
\mathbf{Ric}\equiv \lim_{\gamma\rightarrow0}\,^\gamma\mathbf{Ric}=
\Lambda \,\mathbf{g} + 2\beta\,\delta_0\left(-\mathbf{d}t\,
\mathbf{d}t + 3\mathbf{d}x\, \mathbf{d}x + \mathbf{d}y\,
\mathbf{d}y + \mathbf{d}z\, \mathbf{d}z\right),\label{limRgmp}
\end{equation}
\begin{equation}
\mathbf{G} + \mathbf{g}\,\Lambda \equiv
\lim_{\gamma\rightarrow0}(^\gamma\mathbf{G}+
\,^{\gamma}\mathbf{g}\,\Lambda )=
4\beta\,\delta_0\left(\mathbf{d}t\, \mathbf{d}t - \mathbf{d}y\,
\mathbf{d}y - \mathbf{d}z\, \mathbf{d}z\right),\label{limGgmp}
\end{equation}
where $\mathbf{Ric}$ and $\mathbf{G}$ are the distribution-valued
Ricci and Einstein curvatures of $\mathbf{g}$ \cite{guerrero}. The
diagram (\ref{diagram1}), in which $^\gamma\mathbf{g}$ and
$\mathbf{g}$ are given by (\ref{gmp}) and (\ref{limgmp}),
respectively, holds in the sense of distributions.

Note that $\mathbf{g}$ is piecewise smooth and that the pullback
of $\mathbf{g}$ to the surface $x=0$ is the same from both sides
of this surface. Note also that the six vector fields
(\ref{gmpcoll1}-\ref{gmpcoll2}) are also Killing vectors on the
$(2+1)$-dimensional Minkowski spacetime
$(\mathbb{R}^3,\bar{\mathbf{g}})$ with $\bar{\mathbf{g}}$ the
Minkowski metric which appears as this pullback.

Next, let $\boldsymbol{\eta}$ be the Minkowski metric tensor
(\ref{minkowski}) and let $\nabla$ be the derivative operator in
$\boldsymbol{\eta}$. Let ${\bf U}$ be a test $2$-tensor field on
$\mathbb{R}^4$. From (\ref{liederiv}), it is straightforward to
verify that for the vector fields given by
(\ref{gmpcoll1},\ref{gmpcoll2}), we have
\begin{equation}
\mathcal{L}_{_\mathbf{V}} \,\mathbf{^{\gamma}g}[{\bf
U}]=0\,,\qquad \mathcal{L}_{_\mathbf{V}} \,\mathbf{g}[{\bf U}]=0\,
, \label{liederivgmp}
\end{equation}
as expected. Hence, (\ref{gmpcoll1},\ref{gmpcoll2}) are Killing
vectors of $(\mathbb{R}^4,\mathbf{g})$ in the sense that
$\mathcal{L}_{_\mathbf{V}}\,\mathbf{g}$ along these vectors is the
zero distribution on $\mathbb{R}^4$. Analogous computations show
that
\begin{equation}
\mathcal{L}_{_\mathbf{V}} \,\mathbf{^{\gamma}Ric}[{\bf U}]=0\,
,\qquad \mathcal{L}_{_\mathbf{V}} \,\mathbf{^{\gamma}G}[{\bf
U}]=0\,\label{liederivgmpcurv1}
\end{equation}
and
\begin{equation}
\mathcal{L}_{_\mathbf{V}} \,\mathbf{Ric}[{\bf U}]=0\, ,\qquad
\mathcal{L}_{_\mathbf{V}} \,\mathbf{G}[{\bf U}]=0\,.
\label{liederivgmpcurv2}
\end{equation}
It follows that for this spacetime the diagram (\ref{diagram2})
holds in the sense of distributions. Besides
(\ref{gmpcoll1},\ref{gmpcoll2}), there is no other $C^{\infty}$
vector field for which (\ref{liederivgmpcurv1},
\ref{liederivgmpcurv2}) are satisfied on $\mathbb{R}^4$.

Next, let $\mathcal{M}^+\equiv \{(t,x,y,z)\in \mathbb{R}^4,
\,x>0\}$ and let $\mathbf{U}^{+}$ be a test tensor of compact
support $\mathcal{K}\subset\mathcal{M}^+$. From
(\ref{liederivsmooth}) we find that
\begin{equation}
\mathcal{L}_{_\mathbf{V}}\mathbf{g}[\mathbf{U}^+]=0\label{rightK}
\end{equation}
is satisfied along the $C^{\infty}$ vector fields on
$\mathbb{R}^4$ given by
\begin{eqnarray}
\mathbf{V}_7&=& \boldsymbol{\partial}_x + \beta
(t\boldsymbol{\partial}_t +
y\boldsymbol{\partial}_y + z\boldsymbol{\partial}_z),\label{V7}\\
\mathbf{V}_8&=& \beta t\mathbf{V}_7 +
\left(\frac{1}{2}\beta^2(-t^2+y^2+z^2)+\frac{1}{8}\,
e^{2\beta x}\right)\boldsymbol{\partial}_t,\label{V8}\\
\mathbf{V}_9&=&\beta y\mathbf{V}_7 -
\left(\frac{1}{2}\beta^2(-t^2+y^2+z^2)+\frac{1}{8}\,
e^{2\beta x}\right)\boldsymbol{\partial}_y,\label{V9}\\
\mathbf{V}_{10}&=& \beta z\mathbf{V}_7 -
\left(\frac{1}{2}\beta^2(-t^2+y^2+z^2)+\frac{1}{8}\, e^{2\beta
x}\right)\boldsymbol{\partial}_z. \label{V10}
\end{eqnarray}
It follows that $\mathcal{L}_{_\mathbf{V}}\mathbf{g}$ is the zero
distribution on $\mathcal{M}^+$, along these vector fields.

Now, let $\mathbf{U}$ be a test tensor on $\mathbb{R}^4$ and let
us consider the Lie derivatives (in the sense of distributions) of
(\ref{gmp}) and (\ref{limgmp}) along the vector fields
(\ref{V7}-\ref{V10}). Since $^\gamma\mathbf{g}$ and
$\nabla\,^\gamma\mathbf{g}$ converge locally to $\mathbf{g}$ and
$\nabla\mathbf{g}$ for $\gamma\rightarrow 0$ \cite{guerrero}, it
follows directly that
$\mathcal{L}_{_\mathbf{V}}\,^\gamma\mathbf{g}$ also converge
locally to $\mathcal{L}_{_\mathbf{V}}\mathbf{g}$ along the smooth
vector fields (\ref{V7}-\ref{V10}). Thus, for $\mathbf{V}_7$ we
have
\begin{equation}
\mathcal{L}_{_\mathbf{V_7}}\,^{\gamma}\mathbf{g}=2\beta\cosh^{-2\gamma}\frac{\beta
x}{\gamma} (1-\tanh\frac{\beta x}{\gamma})(-\mathbf{d}t\,
\mathbf{d}t + \mathbf{d}y\, \mathbf{d}y + \mathbf{d}z\,
\mathbf{d}z),\label{nokill-gmp}
\end{equation}
\begin{equation}
\mathcal{L}_{_\mathbf{V_7}}\mathbf{g}=\Theta^-_x\,
4\beta\,e^{2\beta x}(-\mathbf{d}t\, \mathbf{d}t + \mathbf{d}y\,
\mathbf{d}y + \mathbf{d}z\, \mathbf{d}z),\label{rkill-limgmp}
\end{equation}
that satisfy
\begin{equation}
\lim_{\gamma\rightarrow
0}\mathcal{L}_{_\mathbf{V_7}}\,^{\gamma}\mathbf{g}[\mathbf{U}]=
\mathcal{L}_{_\mathbf{V_7}}\mathbf{g}[\mathbf{U}].
\end{equation}
The fact that $\mathcal{L}_{_\mathbf{V_7}}\mathbf{g}$ is the zero
distribution on $\mathcal{M}^+$ may be interpreted naturally as a
Killing symmetry of $\mathbf{g}$ on $\mathcal{M}^+$ generated by
$\mathbf{V}_7$. Since this symmetry arises in the limit
$\gamma\rightarrow 0$ of $^{\gamma}\mathbf{g}$, $\mathbf{V}_7$ is
an {\it asymptotic} Killing vector field on
$({\mathbb{R}}^4,\,^\gamma\mathbf{g})$. The same considerations
holds for the other $C^{\infty}$ vector fields
(\ref{V8}-\ref{V10}). Further, since $(\mathbb{R}^4,\mathbf{g})$
is reflection symmetric along the direction perpendicular to the
wall, a symmetry inherited from
$(\mathbb{R}^4,\,^\gamma\mathbf{g})$, the above considerations can
be extended to $\mathcal{M}^-\equiv \{(t,x,y,z)\in \mathbb{R}^4,
\,x<0\}$ under the replacement $x\rightarrow -x$. Indeed, these
results are by no means unexpected, they simply put in a rigorous
setting the emergence of additional symmetries in the
$\gamma\rightarrow 0$ limit of the spacetime
$(\mathbb{R}^4,\,^\gamma\mathbf{g})$, where we have an absolute
control over what is going on.

For the sake of completeness, let us analyze the action of
$\mathcal{L}_{_\mathbf{V}}$ on the distribution-valued curvature
tensors of the metric (\ref{limgmp}) along the vector fields
(\ref{V7}-\ref{V10}). Let us consider again the vector field
$\mathbf{V}_7$. We find
\begin{equation}
\mathcal{L}_{_\mathbf{V_7}}\mathbf{Ric}= 2\beta
(\delta_0^{\prime}-2\beta \delta_0)\left(-\mathbf{d}t\,
\mathbf{d}t + 3\mathbf{d}x\, \mathbf{d}x + \mathbf{d}y\,
\mathbf{d}y + \mathbf{d}z\, \mathbf{d}z\right)
-\Theta^-_x\,12\beta^3\,e^{2\beta x}(-\mathbf{d}t\, \mathbf{d}t +
\mathbf{d}y\, \mathbf{d}y + \mathbf{d}z\, \mathbf{d}z)
\end{equation}
and
\begin{equation}
\mathcal{L}_{_\mathbf{V_7}}(8\pi\mathbf{T})\equiv
\mathcal{L}_{_\mathbf{V_7}}(\mathbf{G}+\Lambda \mathbf{g})= 4\beta
(\delta_0^{\prime}+2\beta \delta_0)\left(-\mathbf{d}t\,
\mathbf{d}t + \mathbf{d}y\, \mathbf{d}y + \mathbf{d}z\,
\mathbf{d}z\right).
\end{equation}
Indeed, $\mathbf{V}_7$ is neither a Ricci collineation nor a
matter collineation and the same conclusion extends to the vector
fields (\ref{V8}-\ref{V10}). In particular, this shows explicitly
that the distributional energy momentum tensor of the brane does
not inherit all the symmetries of the bulk.

\section{An asymmetric domain wall spacetime}\label{GMspacetime}

Let us now to consider the spacetime
$({\mathbb{R}}^4,\,^\gamma\mathbf{g})$ where the $C^{\infty}$
metric tensor $^\gamma\mathbf{g}$ is given by
\begin{equation}
^\gamma\mathbf{g}= \cosh^{-2\gamma/3}\frac{\beta
x}{\gamma}e^{-4\beta x/3}\left(-\mathbf{d}t\, \mathbf{d}t  +
e^{2\beta x}(\mathbf{d}y\, \mathbf{d}y + \mathbf{d}z\,
\mathbf{d}z)\right)+ \cosh^{-2\gamma}\frac{\beta
x}{\gamma}\mathbf{d}x\, \mathbf{d}x,\label{gass}
\end{equation}
with $\beta$ and $\gamma$ constants and $0<\gamma < 1$. This
represents a two-parameter family of plane symmetric static domain
wall spacetimes in which the reflection symmetry along the
direction perpendicular to the wall has been relaxed, being
asymptotically (i.e far away of the wall) flat for $x>0$ and
behaving asymptotically as the Taub spacetime for $x<0$
\cite{gass}. The metric (\ref{gass}) is solution to the coupled
Einstein-scalar field equations (\ref{eqEinstein}-\ref{eqScalar})
with $\Lambda=0$ and
\begin{equation}
\phi= \phi_0 \tan^{-1}(\sinh(\beta x/\gamma),\qquad \phi_0\equiv
\frac{1}{6}\sqrt{3\gamma(1-\gamma) /\pi}
\end{equation}
and
\begin{equation}
V(\phi)= \frac{1}{24\pi}\beta^2\frac{1}{\gamma}
\cos^{2(1-\gamma)}(\phi/\phi_0),
\end{equation}
where $\gamma$ plays the role of the wall's thickness. The
distributional $\gamma \rightarrow 0$ limit of this geometry has
been analyzed in reference \cite{melfo}. It should be noted that
(\ref{gass}) does not inherit the $\mathrm{Z}_2$ symmetry of
$V(\phi)$, a fact that makes very interesting the analysis of the
symmetries of this spacetime.

The Ricci and Einstein tensor fields of (\ref{gass}) are given by
\begin{equation}
^{\gamma}\mathbf{Ric}= \frac{\beta^2}{3\gamma} (\cosh\frac{\beta
x}{\gamma})^{-2(1-2\gamma/3)}\left( -e^{-4\beta x/3}\mathbf{d}t\,
\mathbf{d}t +
 e^{2\beta x/3}(\mathbf{d}y\, \mathbf{d}y + \mathbf{d}z\,
\mathbf{d}z)\right) +
\frac{\beta^2}{3\gamma}(3-2\gamma)\cosh^{-2}\frac{\beta
x}{\gamma}\mathbf{d}x\, \mathbf{d}x,\label{Rgass}
\end{equation}
\begin{equation}
^{\gamma}\mathbf{G}= \frac{\beta^2}{3\gamma} (\cosh\frac{\beta
x}{\gamma})^{-2(1-2\gamma/3)}(2-\gamma)\left( e^{-4\beta
x/3}\mathbf{d}t\, \mathbf{d}t - e^{2\beta x/3}(\mathbf{d}y\,
\mathbf{d}y + \mathbf{d}z\, \mathbf{d}z)\right) -
\frac{\beta^2}{3}\cosh^{-2}\frac{\beta x}{\gamma}\mathbf{d}x\,
\mathbf{d}x,\label{Ggass}
\end{equation}
and the distributional thin wall limits of
(\ref{gass},\ref{Rgass},\ref{Ggass}) are given by \cite{melfo}
\begin{eqnarray}
\mathbf{g}\equiv \lim_{\gamma\rightarrow0}\,^\gamma\mathbf{g}&=&
\Theta^{-}_x\left(-e^{-2\beta x/3}\mathbf{d}t\, \mathbf{d}t +
e^{2\beta x}\mathbf{d}x\, \mathbf{d}x + e^{4\beta
x/3}(\mathbf{d}y\, \mathbf{d}y + \mathbf{d}z\,
\mathbf{d}z)\right)\nonumber\\ &\,&+
\,\Theta^{+}_x\left(-e^{-2\beta x}\mathbf{d}t\, \mathbf{d}t +
e^{-2\beta x}\mathbf{d}x\, \mathbf{d}x + \mathbf{d}y\, \mathbf{d}y
+ \mathbf{d}z\, \mathbf{d}z\right),\label{limgass}
\end{eqnarray}
\begin{equation}
\mathbf{Ric}\equiv \lim_{\gamma\rightarrow0}\,^\gamma\mathbf{Ric}=
\frac{1}{3}\beta\,\delta_0\left(-\mathbf{d}t\, \mathbf{d}t +
3\mathbf{d}x\, \mathbf{d}x + \mathbf{d}y\, \mathbf{d}y +
\mathbf{d}z\, \mathbf{d}z\right),\label{limRgass}
\end{equation}
\begin{equation}
\mathbf{G}\equiv \lim_{\gamma\rightarrow0}\,^\gamma\mathbf{G}=
\frac{2}{3}\beta\,\delta_0\left(\mathbf{d}t\, \mathbf{d}t -
\mathbf{d}y\, \mathbf{d}y - \mathbf{d}z\,
\mathbf{d}z\right).\label{limGgass}
\end{equation}
Note that $\mathbf{g}$ is piecewise smooth and that the pullback
of $\mathbf{g}$ to the surface $x=0$ is the same from both sides
and coincides with the ordinary $(2+1)$-dimensional Minkowski
metric. Like its smoothed version (\ref{gass}), $\mathbf{g}$ is
not reflection symmetric along the coordinate perpendicular to the
wall. The spacetime $(\mathbb{R}^4,\mathbf{g})$ for $x>0$ is
isometric to the ordinary $(3+1)$-dimensional Minkowski spacetime,
while for $x<0$ it is the Taub spacetime \cite{gass,melfo}.

There are only four independent Killing vector fields on
$({\mathbb{R}}^4,\,^\gamma\mathbf{g})$ and these are given by
\begin{equation}
\qquad \mathbf{V}_1= \boldsymbol{\partial}_y,\qquad \mathbf{V}_2=
\boldsymbol{\partial}_z,\qquad \mathbf{V}_3=
z\boldsymbol{\partial}_y -y\boldsymbol{\partial}_z,\qquad
\mathbf{V}_4= \boldsymbol{\partial}_t. \label{gasscoll1}
\end{equation}
Let $\nabla$ be the derivative operator in the ordinary Minkowski
metric tensor (\ref{minkowski}). Let ${\bf U}$ be a test
$2$-tensor field on $\mathbb{R}^4$. From (\ref{liederiv}) it is
straightforward to show that the four vectors given in
(\ref{gasscoll1}) generate also isometries of (\ref{limgass}) and
are improper collineations for the distributional Ricci and
Einstein curvature tensors. It follows that for this spacetime the
diagram (\ref{diagram2}) holds in the sense of distributions.

Next, consider the $C^{\infty}$ vector fields given by
\begin{equation}
\mathbf{V}_5= t\boldsymbol{\partial}_y +
y\boldsymbol{\partial}_t,\qquad \mathbf{V}_6 =
z\boldsymbol{\partial}_t +t\boldsymbol{\partial}_z\, .
\label{gasscoll2}
\end{equation}
We have
\begin{eqnarray}
\mathcal{L}_{_\mathbf{V_5}}\,\mathbf{g}[{\bf U}]&\equiv& -
\mathbf{g}[\mathcal{L}_{_\mathbf{V_5}} \mathbf{U} +
(\nabla\cdot\mathbf{V_5})\mathbf{U}]\nonumber\\&=&
-\int_{\mathbb{R}^4}\,{\bf g}
\cdot(\mathcal{L}_{_\mathbf{V_5}}{\bf U} +
\mathbf{U}(\nabla\cdot\mathbf{V_5}))\boldsymbol{\omega}_{\eta}\nonumber\\&=&
-\int_{x<0}(e^{-2\beta x/3}-e^{4\beta x/3})(U^{ty} +
U^{yt})dt\,dx\,dy\,dz \nonumber\\& \, & -\int_{x>0}(e^{-2\beta
x}-1)(U^{ty} + U^{yt})dt\,dx\,dy\,dz ,
\end{eqnarray}
where we have integrated by parts and used the fact that
$\mathbf{U}$ is of compact support. It follows
\begin{equation}
\mathcal{L}_{_\mathbf{V_5}}\,\mathbf{g}=2\left( \Theta^-_x
e^{\beta x/3} + \Theta^+_x e^{-\beta x} \right) \sinh\beta
x\,(\mathbf{d}t\, \mathbf{d}y + \mathbf{d}y\, \mathbf{d}t ).
\end{equation}
On the other hand, as can be guessed from the explicit form of
$\mathbf{Ric}$, we have
\begin{eqnarray}
\mathcal{L}_{_\mathbf{V_5}}\,\mathbf{Ric}[{\bf U}]&\equiv& -
\mathbf{Ric}[\mathcal{L}_{_\mathbf{V_5}} \mathbf{U} +
(\nabla\cdot\mathbf{V_5})\mathbf{U}]\nonumber\\&=&
-\frac{1}{3}\beta\int_{x=0}(y\partial_t + t\partial_y)(U^{tt} +
3U^{xx} + U^{yy} + U^{zz})\,dt\,dy\,dz \nonumber\\&=&0,
\end{eqnarray}
where in last step we have integrated by parts and used again the
fact that $\mathbf{U}$ is of compact support. Analogous
computations show that
\begin{equation}
\mathcal{L}_{_\mathbf{V_6}}\,\mathbf{g}=2\left( \Theta^-_x
e^{\beta x/3} + \Theta^+_x e^{-\beta x} \right) \sinh\beta
x\,(\mathbf{d}t\, \mathbf{d}z + \mathbf{d}z\, \mathbf{d}t),
\end{equation}
\begin{equation}
\mathcal{L}_{_\mathbf{V_6}}\,\mathbf{Ric}=0
\end{equation}
Hence, although $V_5$ and $V_6$ are not Killing vectors of
$(\mathbb{R}^4,\mathbf{g})$, they are Ricci collineations.
Further,
\begin{equation}
\mathcal{L}_{_\mathbf{V_5}}\,\mathbf{G}=0, \qquad
\mathcal{L}_{_\mathbf{V_6}}\,\mathbf{G}=0.
\end{equation}
and $V_5$ and $V_6$ are also Einstein collineations. Indeed, the
above expressions should be understood in the sense of
distributions. Besides (\ref{gasscoll1},\ref{gasscoll2}), there is
no other $C^{\infty}$ vector field that generates symmetries of
$\mathbf{Ric}$ and $\mathbf{G}$.

It should be noted that $\mathcal{L}_{_\mathbf{V_5}}\,\mathbf{g}$
and $\mathcal{L}_{_\mathbf{V_6}}\,\mathbf{g}$ are piecewise smooth
and that the pullbacks of
$\mathcal{L}_{_\mathbf{V_5}}\,\mathbf{g}$ and
$\mathcal{L}_{_\mathbf{V_6}}\,\mathbf{g}$ to the surface $x=0$ are
the same from both sides of the wall and vanish. It follows that
the symmetries of the distribution-valued curvature tensors
$\mathbf{Ric}$ and $\mathbf{G}$ (which are supported on the
surface $x=0$) coincide with the Killing symmetries of the
pullback of $\mathbf{g}$ to this surface. For the sake of
completeness, let us show that the same result is also reached
within the standard thin shell formalism \cite{israel,barrabes}.
For $(\mathbb{R}^4,\mathbf{g})$ with $\mathbf{g}$ given by
(\ref{limgass}), the pullback $\bar{\mathbf{g}}$ of $\mathbf{g}$
to the surface $x=0$ is the same from both sides and is given by
\begin{equation}
\bar{\mathbf{g}}=\left(-\mathbf{d}t\, \mathbf{d}t + \mathbf{d}y\,
\mathbf{d}y + \mathbf{d}z\, \mathbf{d}z\right).
\end{equation}
The extrinsic curvature tensor $\mathbf{K}$ of the surfaces
$x=x_0$ as submanifolds of $(\mathbb{R}^4,\mathbf{g})$, is the
$C^{\infty}$ regularly discontinuous tensor across the surface
$x=0$ given by
\begin{equation}
\mathbf{K}= -\frac{1}{3}\beta\Theta^-_x \left(e^{-2\beta
x_0/3}\mathbf{d}t\, \mathbf{d}t + 2e^{-4\beta x_0/3}(\mathbf{d}y\,
\mathbf{d}y + \mathbf{d}z\, \mathbf{d}z)\right) - \beta \Theta^+_x
e^{-2\beta x_0}\mathbf{d}t\, \mathbf{d}t\, .
\end{equation}
Thus, the discontinuity $[[ \mathbf{K} ]]\equiv
\mathbf{K}|_{x_0=0^+} - \mathbf{K}|_{x_0=0^-}$ of $\mathbf{K}$
across the surface $x=0$, which is declared as a purely intrinsic
property of this surface, is the ordinary $C^{\infty}$ tensor
field defined on the surface $x=0$ given by
\begin{equation}
[[ \mathbf{K} ]]=\frac{2}{3}\beta\left(-\mathbf{d}t\, \mathbf{d}t
+ \mathbf{d}y\, \mathbf{d}y + \mathbf{d}z\, \mathbf{d}z\right).
\end{equation}
Therefore, the symmetries of $[[\mathbf{K}]]$ are the symmetries
of the pullback $\bar{\mathbf{g}}$ of $\mathbf{g}$ to the surface
$x=0$. Now, we have
\begin{equation}
\mathbf{Ric}=\frac{1}{2}\delta_0\left( \,[[\mathbf{K}]] +
(\mathbf{\bar{\mathbf{g}}}^{-1}\cdot[[\mathbf{K}]])\,
{\mathbf{d}}x
\,{\mathbf{d}}x\,\right)\,,\label{limRgass-extrinsic}
\end{equation}
from which it follows that the symmetries of the distributional
Ricci tensor of $\mathbf{g}$ turns out to be the Killing
symmetries of the induced metric $\bar{\mathbf{g}}$ on the surface
$x=0$ where the $2$-brane is localized. It should be noted that
this is not an example of the well known (trivial) symmetry
inheritance. Although, the pullback $\bar{\mathbf{g}}$ of
$\mathbf{g}$ to the surface $x=0$ acts a metric on this surface
(as follows from the fact that the surface $x=0$ has a
well-defined intrinsic geometry) and therefore the Killing vectors
of this pullback are naturally collineations of the curvature
tensors of this pullback, $\mathbf{Ric}$ given by (\ref{limRgass})
or (\ref{limRgass-extrinsic}) is not the Ricci tensor of
$\bar{\mathbf{g}}$. The same conclusions can be extended also to
all the thin domain wall geometries considered in the previous
sections.

Remarkably enough, the asymmetric geometry considered here,
explicitly shows that the Killing symmetries of the pullback of
the metric tensor to the surface where the thin wall is localized
may form a larger group than the group of Killing symmetries which
are common to the ambient spacetime on both sides of the wall.

\section{Summary and Discussion}

In this work, by generalizing the Lie derivative of smooth tensor
fields to distribution-valued tensors, we defined and computed the
Killing symmetries and the Ricci and Einstein collineations of
some distributional domain wall geometries for which the diagram
(\ref{diagram1}), with $^\gamma\mathbf{g}$ and $\mathbf{g}$ the
distribution valued metric tensors of the thick and thin domain
wall spacetimes respectively, holds rigourously in the sense of
distributions. For all the geometries considered, the
distribution-valued curvature tensors of the thin wall limit have
singular parts proportional to a Dirac distribution supported on
the surface $\Sigma$ where the thin wall is localized. We found
that the Killing symmetries of the distributional geometry of the
thin wall spacetime $(\mathbb{R}^4,\mathbf{g})$ are the Killing
symmetries of the smooth thick domain wall spacetime
$(\mathbb{R}^4,\,^{\gamma}\mathbf{g})$ and that, besides these,
there are no other isometries. However, as expected, the thin wall
geometry may shows additional symmetries on the open disjoint sets
$\mathcal{M}^+$ and $\mathcal{M}^-$, that admit $\Sigma$ as a
boundary, which are not isometries inherited from the
corresponding smooth geometry. These symmetries are the asymptotic
(i.e. far away of the wall) Killing symmetries of
$(\mathbb{R}^4,\,^{\gamma}\mathbf{g})$.

For the thin domain walls with reflection symmetry of sections
\ref{Gspacetime} and \ref{GMPspacetime}, the Killing vectors of
$(\mathbb{R}^4,\mathbf{g})$ are the only symmetries of the
corresponding distribution-valued Ricci and Einstein tensors.
Therefore the Ricci and Einstein collineations of these thin wall
geometries are improper. For the asymmetric thin domain wall of
section \ref{GMspacetime}, we found that the collineations of the
distributional Ricci and Einstein tensors form a larger group than
the one formed by the Killing symmetries of the corresponding
spacetime $(\mathbb{R}^4,\mathbf{g})$. The additional symmetries
are then proper Ricci and Einstein collineations. Finally, for all
the thin wall spacetimes $(\mathbb{R}^4,\mathbf{g})$ considered,
the symmetries of the distributional curvature tensors turns out
to be the Killing symmetries of $(\Sigma,\bar{\mathbf{g}})$, where
$\bar{\mathbf{g}}$ is the pullback of $\mathbf{g}$ to $\Sigma$.

Although we have restricted ourselves to consider four-dimensional
domain wall spacetimes, these models are straightforwardly
generalized to $\mathrm{D}$-dimensional domain wall spacetimes
\cite{melfo}. On the other hand, the analysis presented here can
be carried out, in principle, for all the distribution-valued
curvature tensors of a spacetime $(\mathcal{M},\mathbf{g})$,
whenever the distribution-valued metric tensor $\mathbf{g}$ is a
regular metric in the sense of reference \cite{geroch}. This
generalization and its implications will be discussed in a
forthcoming paper.

\acknowledgments We wish to thank A. Melfo, L. A. Nu{\~n}ez, Y. Parra
and U. Percoco for enlightening discussions.


\begin{thebibliography}{999}

\bibitem{wald} R. Wald, {\it General Relativity} (The University
 of Chicago Press, 1984).

\bibitem{katzin} G. H. Katzin, J. Levine and W. R. Davis,
 {\it J. Math. Phys.} {\bf 10} (1969) 617.
\bibitem{coley} A. A. Coley and B. O. J. Tupper, {\it J. Math. Phys.}
 {\bf 30} (1989) 2616.
\bibitem{carot} J. Carot, J. da Costa and E. G. L. R. Vaz, {\it J. Math. Phys.}
 {\bf 35} (1994) 4832.
\bibitem{contreras} G. Contreras, L. A. Nu{\~n}ez and U. Percoco, {\it
 Gen. Rel. Grav.} {\bf 32} (2000) 285, \texttt{gr-qc/9907075}.

\bibitem{geroch} R. P. Geroch and J. Traschen, {\it Phys. Rev. D}
 {\bf 36} (1987) 1017.
\bibitem{israel} W. Israel, {\it Nuovo Cimento} {\bf 44} (1966) 1;
 {\it ibid} {\bf 48} (1967) 463.
\bibitem{khorrami} R. Mansouri and M. Khorrami, {\it J. Math. Phys.}
 {\bf 37} (1996) 5672, \texttt{gr-qc/9608029}.

\bibitem{aichelburg94} P. C. Aichelburg and H. Balasin, {\it Class. Quantum Grav.}
 {\bf 11} (1994) L121, \texttt{gr-qc/9407018}.
\bibitem{aichelburg71} P. C. Aichelburg and R. U. Sexl, {\it Gen. Rel.
 Grav.} {\bf 2} (1971) 303.
\bibitem{aichelburg95} P. C. Aichelburg and H. Balasin, {\it Class. Quantum Grav.}
 {\bf 13} (1996) 723, \texttt{gr-qc/9509025}.
\bibitem{aichelburg96} P. C. Aichelburg and H. Balasin, {\it Class. Quantum Grav.}
 {\bf 14} (1997) A31, \texttt{gr-qc/9607045}.

\bibitem{randall} L. Randall and R. Sundrum, {\it Phys. Rev. Lett.}
 {\bf 83} (1999) 4690, \texttt{hep-th/9906064}.


\bibitem{vilenkin} A. Vilenkin, {\it Phys. Lett.}
 {\bf B133} (1983) 177.

\bibitem{cvetic93} M. Cvetic, S. Griffies and H. H. Soleng, {\it Phys. Rev.}
 {\bf D48} (1993) 2613, \texttt{gr-qc/9306005}.
\bibitem{cvetic97} M. Cvetic H. H. Soleng, {\it Phys. Rept.}
 {\bf 282} (1997) 159, \texttt{hep-th/9604090}.

\bibitem{goetz} G. Goetz, {\it J. Math. Phys.} {\bf 31} (1990) 2683.
\bibitem{mukherjee} M. Mukherjee, {\it Class. Quantum Grav.}
 {\bf 10} (1993) 131.
\bibitem{bonjour} F. Bonjour, C. Charmousis and R. Gregory, {\it Class. Quantum Grav.}
 {\bf 16} (1999) 2427, \texttt{gr-qc/9902081}.
\bibitem{gass} R. Gass and M. Mukherjee, {\it Phys. Rev.} {\bf D60} (1999) 065011,
 \texttt{gr-qc/9903012}.

\bibitem{dewolfe} O. deWolfe, D. Z. Freedman, S. S. Gubser and A.
 Karch, {\it Phys. Rev.} {\bf D62} (2000) 046008, \texttt{hep-th/9909134}.
\bibitem{gremm} M. Gremm, {\it Phys. Lett.} {\bf B478} (2000) 434,
 \texttt{hep-th/9912060}.
\bibitem{wang} A. Wang, {\it Phys. Rev.} {\bf D66} (2002) 024024,
 \texttt{hep-th/0201051}.

\bibitem{guerrero} R. Guerrero, A. Melfo and N. Pantoja, {\it Phys. Rev.}
 {\bf D65} (2002) 125010, \texttt{gr-qc/0202011}.
\bibitem{melfo} A. Melfo, N. Pantoja and A. Skirzewski, {\it Phys. Rev.}
 {\bf D67} (2003) 105003, \texttt{gr-qc/0211081}.

\bibitem{choquet} Y. Choquet-Bruhat and C. DeWitt-Morette with M.
 Dillard-Bleick, {\it Analysis, Manifolds and Physics. Part I: Basics.}
 (North-Holland, 1991).
\bibitem{choquet2} Y. Choquet-Bruhat and C. DeWitt-Morette,
{\it Analysis, Manifolds and Physics. Part II.} (North-Holland,
2000).
\bibitem{lichnerowicz} A. Lichnerowicz, {\it Propagateurs, Commutateurs
 et Anticommutateurs en Relativit\'{e} G\'{e}n\'{e}rale}, Publication I.H.E.S.
 No. 10 (1961).
\bibitem{dray} T. Dray, {\it Int. J. Mod. Phys.} {\bf D6} (1997)
717; \texttt{gr-qc/9701047}.
\bibitem{rham} G. de Rham, {\it Differentiable Manifolds} (Springer
 Verlag, 1984).

\bibitem{barrabes} C. Barrabes and W. Israel, {\it Phys. Rev.}
 {\bf D43} (1991) 1129.


\end{thebibliography}
\end{document}